# Conditions for thermoelectric power factor improvements upon band alignment in complex bandstructure materials

Saff E Awal Akhtar[+] and Neophytos Neophytou[*]

School of Engineering, University of Warwick, Coventry, CV4 7AL, UK

[+]saff-e-awal.akhtar@warwick.ac.uk, [*]N.Neophytou@warwick.ac.uk

## Abstract

Band alignment (or band convergence) is a strategy suggested to provide improvements in the thermoelectric power factor (PF) of materials with complex bandstructures. The addition of more bands at the energy region that contributes to transport, can provide more conducting paths and could improve the electrical conductivity and PF of a material. However, this can lead to increased inter-valley scattering, which will tend to degrade the conductivity. Using the Boltzmann Transport Equation (BTE) and a multi-band model, we theoretically investigate the conditions under which band alignment can improve the PF. We show that PF improvements are realized when intra-band scattering between the aligned bands dominates over inter-band scattering, with larger improvements reached when a light-band is brought into alignment. In the more realistic scenario of intra-and inter-band scattering co-existence, we show that in the light band alignment case, possibilities of PF improvement are present even down to the level where the intra- and inter-band scattering are of similar strength. For heavy band alignment this tolerance is weaker, and weaker inter-band scattering is necessary to realize PF improvements. On the other hand, when inter-band scattering dominates, it is not possible to realize any PF improvements upon band alignment, irrespective of bringing a light or a heavy band into alignment. Overall, to realize PF improvements upon band alignment, the valleys that are brought into alignment need to be as electrically conducting as possible compared to the lower energy base valleys and interact as little as possible with those.

Index terms: thermoelectric materials, theory and simulation, thermoelectric power factor, multi-band model, band alignment, band convergence.

# I. Introduction

Thermoelectric (TE) materials convert heat into electricity and vice versa and could be used for power generation as well as cooling applications [1-6]. As such, they could contribute to energy sustainability and reduction of the use of fossil fuels. Their efficiency is quantified by the dimensionless figure of merit, $ZT = \sigma S^2 T/\kappa$, where $S$ stands for the Seebeck coefficient, $\sigma$ for the electrical conductivity, $T$ for the absolute temperature, and $\kappa$ for the thermal conductivity. The product $\sigma S^2$ is called the power factor (PF) and quantifies the power production of the process. For the last two decades, the focus of the TE research targeted thermal conductivity reductions, and with the successful implementation of nanostructuring, large $ZT$ improvements have been reached [4-10]. The $ZT$ has reached values of over 2 in many material cases and temperatures, doubling its values from around unity for the handful of materials of interest ($Bi_2Te_3$, PbTe, SiGe) as of a decade ago [11-14]. These achievements expanded the material space exploration to many, more abundant, inexpensive, and environment friendly materials such as Half-Heusler alloys [15-20], selenides and tellurides [21-24], clathrates [25, 26], skutterudites [27-30], etc.

Similar order of benefits from the PF, however, were not achieved, because the Seebeck coefficient ($S$) and electrical conductivity ($\sigma$) are adversely inter-connected, and optimising one frequently has a negative impact on the other. On the other hand, novel approaches related to band engineering have allowed for some PF improvements. With bandstructure engineering one alters the electronic structure of the material typically through alloying, to realize ideal circumstances for a high Seebeck coefficient and electrical conductivity, an effort to relax the adverse interdependence of the two quantities. Typically, modern TE materials have rich electronic structures, which consist of many bands and valleys, of varying degeneracies and effective masses, which allow for such design approaches [31-35]. The most followed direction for PF improvement is band alignment (also referred to as band convergence). This refers to shifting the relative energy locations of the different valleys of the different bands to achieve energy alignment, thus increasing the number of channels available for transport [36-40]. It can prove beneficial in two ways: the many aligned valleys transport channels can improve the carrier density and conductivity, but they also increase the density of states at the band edge which increases the Seebeck coefficient, both of which can lead to PF improvements [37,41-44].

The complexity of the electronic structure can offer such PF possibilities; however, complex electronic structures also involve complex electronic scattering physics, some of which can be detrimental to the PF [35,45-49]. An important scattering aspect is inter-band/valley scattering, under which carriers from one band/valley can scatter into another (into a band/valley that is brought into alignment, for example). In this case, although the number of transport bands/valleys increases, carrier scattering can also increase, and the benefits of alignment are reduced [37,45,50]. Beyond scattering details, prior computational studies have also suggested that the attributes of the initial lower energy valley (base valley) with respect to the valley(s) that are brought into alignment (aligned valleys), also matter, with more benefits realized when light aligned valleys are introduced [51,52]. On the other hand, the base and aligned valleys also have in general different degeneracies and different scattering details in between themselves as well. Thus, in order to have *a priori* indication if a given electronic structure will potentially lead to PF improvements upon band alignment through alloying or strain processes, the following questions need to be answered: i) at what degree does the interplay between intra- versus inter-valley scattering allows for PF improvements? ii) what ratios between the effective masses of the base and aligned valleys are beneficial? iii) how does the degree of valley degeneracy of the base or the aligned bands affect PF improvements?

In this paper, we employ a multi-band parabolic model and the Boltzmann Transport Equation (BTE) including energy dependent scattering rates, with the goal of answering the questions posed above. We aim to provide a full investigation of the bandstructure and scattering parameter landscape that allows for PF improvements upon band alignment in complex bandstructure TE materials. We show that in general, PF improvements can only be realized when: i) intra-valley scattering dominates over inter-valley scattering, and ii) those benefits are larger when the aligned valley has a lighter effective mass. We also show that in the best case of realistic scenarios, such benefits can only lead to around doubling the PF, which is still important. In general, we suggest that for PF improvements, the aligned valleys need to be as conducting as possible compared to the base valleys, and interact as little as possible between themselves, and between the base valleys. Our results narrow down the design window for PF improvements, which would be highly beneficial for aiding experimentalists to target more effectively materials that can actually provide improvements. On a more optimistic note, it is suggested that some of the best new generation TE materials are polar optical phonon scattering limited or ionized impurity scattering limited [49,53]. Since these are primarily anisotropic, intra-valley scattering mechanisms, these stringent criteria are somewhat weakened.

The paper is organized as follows. In Section II we describe the methodology used to calculate the PF using a parabolic multi-band model within the Boltzmann Transport Equation (BTE) formalism. In Section III we present our results and discuss the case of band alignment of single valley band under different scattering conditions. In Section IV we present our results for band alignment of multi-valley bands. Finally, in Section V we conclude.

## II. Methodology

We have used a multi-band parabolic effective mass model to study the effect of band alignment under different scattering considerations. The Boltzmann transport equation (BTE) theory under the energy dependent relaxation time approximation was used to evaluate the transport coefficients. The thermoelectric coefficients i.e. electrical conductivity ($\sigma$), Seebeck coefficient (*S*), and power factor (PF) at a particular Fermi level, $E_F$, are computed as [54,55]:

$$\sigma_{(E_F)} = q^2 \int_E G(E) \left(-\frac{\partial f}{\partial E}\right) dE \tag{1}$$

$$S_{(E_F)} = \frac{qk_B}{\sigma_{(E_F)}} \int_E G(E) \left(\frac{E-E_F}{k_B T}\right) \left(-\frac{\partial f}{\partial E}\right) dE \tag{2}$$

$$\text{PF} = \sigma_{(E_F)} S_{(E_F)}^2 \tag{3}$$

Above $q$ is the electronic charge, $f$ is the fermi distribution function, $k_B$ is the Planck's constant, $E_F$ is the Fermi energy level, and $G(E)$ is the transport distribution function (TDF), defined as:

$$G(E) = \tau_s(E) v^2(E) g(E) \tag{4}$$

where $v$ is the bandstructure velocity, $g$ is the density of states, and $\tau_s$ is the scattering (momentum) relaxation time. The band velocity and density of states are directly extracted from the $E(k)$ relationship, $E(k) = \frac{\hbar^2 k^2}{2m_o^*}$ as:

$$v_n(E) = (2E/m_n^*)^{1/2} \tag{5}$$

$$g(E) = 2^{1/2} m_n^{*3/2} N_n E^{1/2}/(\pi^2 \hbar^3) \tag{6}$$

where $m_n^*$ is the mass, and $N_n$ is the the degeneracy of the $n^{th}$ band. The scattering time in Eq. (4) is given by [54]:

$$\frac{1}{\tau_{nm}(E)} = \frac{\pi D_{nm}^2}{2\rho \omega_{ph}} \left( N_\omega + \frac{1}{2} \mp \frac{1}{2} \right) g_{s,m}(E \pm \hbar \omega_{ph}) \tag{7}$$

where we consider a single electron-optical phonon process, which typically play dominat part in inter-valley processes. These are are central in our analysis below (by being able to support transitions of large energy and momentum exchange), but without loss in generality. Above, $g_{s,m}$ is the scattering density of states (that carriers scatter into) for absorption and emission, $N_\omega$ is the Bose-Einstein distribution function, and $\omega_{ph}$ is the frequency of the phonons, for which we use $\omega_{ph} = 0.03 eV$, typical for TE materials i.e half-Heusler alloys [50]. $D_{nm}$ is the inelastic optical deformation potential that defines the optical phonon scattering strength. In this work we use values from $1.2 \times 10^{10}$ (eV/m) to $12 \times 10^{10}$ (eV/m), again typical for semiconducotrs and TE materials [50,53]. The subscripts $n$ and $m$ correspond to the different bands that interrract through the scatering process. We consider both, intra-band/valley, as well as inter-band/valley transitions, which can be facilitated by different deformation potentials. In the first case, $m = n$ and it results to $\tau_{nn}$ with $D_{nn}$ (which we will refer to as $D_{intra}$), where in the second case $m \neq n$, which results to $\tau_{nm}$ with $D_{nm}$ (which we will refer to as $D_{inter}$). The overall scattering rate from an initial state at energy $E$ and band/valley $n$ is then computed using Matthiessen's rule as:

$$\frac{1}{\tau_{n,total}(E)} = \frac{1}{\tau_{nn}(E)} + \frac{1}{\tau_{nm}(E)} \tag{8}$$

In Fig. 1 we provide an illustrative example of the studies and analysis that follows. In the presence of intra/inter-band scattering, full band alignment of two valleys that are initially separated in energy affects the TE coefficients by increasing the number of conducting channels, but also the scattering states. We label the lower energy band as the 'Base-B' band and the aligned band as the 'Aligned-A' band. Figure 1a-c (left column) shows the effect on the TE coefficients (electrical conductivity, Seebeck coefficient and PF) upon *aligning a light band*. As the band aligns from $\Delta E = 0.2$eV down to $\Delta E = 0$eV, the electrical conductivity slightly increases due to the more conducting light band that now participates in transport, whereas the Seebeck coefficient remains unaffected since it is determined at first order by the band edge of the lowest band. Thus, a 33% PF improvement is observed. However, *aligning a heavy band*, as shown in Fig. 1d-f (right column), is detrimental to the PF with a 55% reduction upon full band alignment, due to increase in scattering that the base band carriers experience into the large DOS of the heavy/ low-conducting aligned band. Thus, band alignment is not always beneficial, and in fact in the latter case it should be avoided [48,51]. Below we present a full analysis of when band alignment can, and when it cannot be beneficial.

Note that in both cases, the dominant factor that determines the change in the PF is the electronic conductivity, whereas the Seebeck coefficient has less of an effect. In fact, as shown in the insets in Fig. 1, the Seebeck coefficient as expected follows the inverse trend compared to conductivity, and consequently the inverse trend compared to the PF as well. Also note that in our data for the Seebeck coefficient in Fig. 1, the Seebeck coefficient only slightly shifts and that is only observed at high $E_F$. Typically, one might expect that once an additional band is aligned, and the density of states (DOS) at the band edge increases, it will result in an increase in the Seebeck coefficient. Note, however, that this behavior is only observed when the Seebeck coefficients are plotted as a function of density. We present this in Fig S4.1 and Fig S4.2 in the Supporting Information file together with appropriate discussions. We show that, clearly, if plotted as a function of density, the entire Seebeck coefficient line, upon full band alignment acquires a right shift, indicating an increase in DOS (in a similar way to the typical interpretation of right shifts in the Pisarenko plot as increase in the effective mass). This is evident more in the case of heavy aligned band which increases the DOS substantially, compared to the case of the light aligned mass which does not cause significant changes in the DOS. On the other hand, PF improvements are reached in the latter case of light aligned band, clearly indicating that the conductivity is that determines the improvements. Also note that the conductivity will experience the same shift when plotted versus density, such that the PF (and its maximum) does not change, as further discussed in the Supporting Information file.

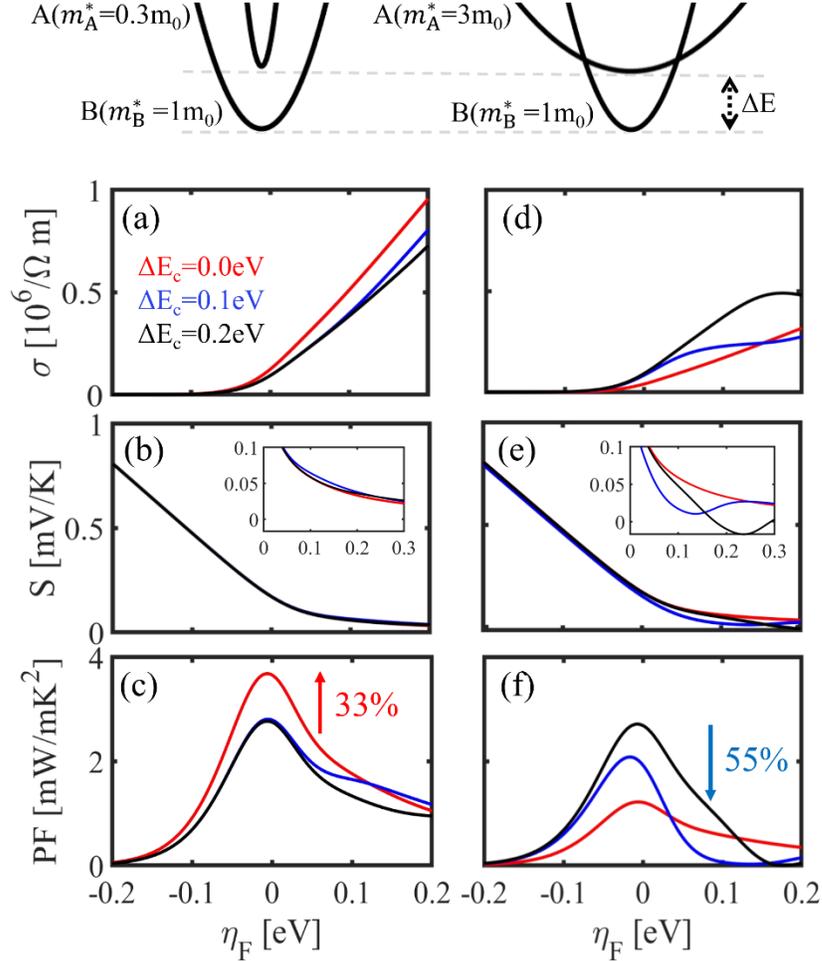

**Figure 1:** Band alignment illustrated for: aligning a light effective mass band (left column) and aligning a heavy effective mass band (right column), as shown in the top schematics. The corresponding transport coefficients, conductivity, Seebeck coefficient and PF versus Fermi level, $E_F$ are indicated in each column, (a-c) and (d-f) respectively. Here $m_B^*$ is the effective mass of the base band (B) while $m_A^*$ is the effective mass of the aligned band (A) in the unit of $m_0$. The PF percentage change upon band alignment for the two cases is noted. The insets show a zoom-in of the Seebeck coefficient at high $E_F$, where the PF peaks. We set $E = 0$ eV and $\eta_F = 0$ eV at the base band edge.

An interesting observation is the trend of the Seebeck coefficient at high $E_F$ for the heavy aligned band case (inset of Fig. 1b). The fully aligned red line is the highest, whereas for the misaligned cases the Seebeck is lower. A cross-over in the cases of $\Delta E = 0.2$ eV (black line) and $\Delta E = 0.1$ eV (blue line) is observed, which can be explained as follows: Starting from the misaligned case (black line) the Seebeck coefficient is low. There is a noticeable reduction at $E = 0.2$ eV, where the upper band is met. The reason is that the upper band is of high mass and low conductivity, with states of reduced current capabilities. Thus, the average energy of the current flow (which defines the Seebeck coefficient – see discussions in the Supporting Information file) is shifted to lower energies and the Seebeck is reduced, even becoming negative for some interval (indicating that transport below the Fermi level is stronger). Note that although we consider electrons with negative Seebeck, in all cases we consider in this work we flip the sign of the Seebeck coefficient for convenience. Similarly in the case of the blue line with partial alignment, this deep is shifted at the energy where the upper band is encountered at $E = 0.1$ eV. At higher energies the Seebeck coefficient of the two misaligned cases will tend to approach that of the fully aligned case, and thus the crossover of the two misaligned cases. The fully aligned band case has the highest Seebeck coefficient, as the average energy of the current flow is not hindered by scattering from heavier upper valleys. Notice that, clearly, the Seebeck coefficients follow the expected inverse trend compared to the conductivity.

## III. Band alignment with single-valley bands

Typically, the complex electronic structure in materials consists of several bands and valleys, which can differ arbitrarily in energy, and consist of arbitrarily different effective masses. The scattering processes are also complex in general, with different scattering strengths (and deformation potentials) determining the intra-band/valley processes for each band, and the inter-band processes, which can be of different strength as well [34,53]. Thus, the parameter space for exploration of when band alignment is beneficial is large.

To simplify our investigation and build a first order understanding of the influence of band alignment on the TE properties, we begin by studying a simple system of two single-valley bands, separated by energy $\Delta E$. We then assume a common intra-band deformation potential, $D_{intra}$, but allow for different inter-band deformation potential, $D_{inter}$. This will allow us to focus on the effect of intra- versus inter-band scattering, i.e. control the degree of scattering as bands are aligned. We consider a base band ('B') of effective mass $m_B^* = 1m_0$ and a band that we bring into alignment (referred to as the aligned band – 'A'), for which its mass $m_A^*$ is either light or heavy. Distinguishing between the *'light band alignment'* versus the *'heavy band alignment'* cases is also essential in identifying promising design cases, as indicated in Fig. 1. In the investigations from here on, for each considered case, we compute the TE power factor as a function of the reduced Fermi level $\eta_F = E_F - E_C$ as in Fig. 1, and extract the peak of that PF. We then plot that peak value as a function of the energy separation between the bands, $\Delta E$, and for the parameters under investigation ($D_{intra}$, $D_{inter}$, $m_A^*$, $m_B^*$).

*Intra- versus inter-band scattering for light and heavy band alignment:* As a first study we fix and set equal values of $D_{intra}$ and $D_{inter}$ and focus on the effect of aligning bands of different masses, as well as the effect of intra- versus inter-band scattering between the base and aligned bands. The schematics in the left column of Fig. 2 (Fig. 2a and Fig. 2d) illustrate the distinct cases considered. In Fig. 2a-c we consider aligning a light band to the base band. Here we use $m_A^* = 0.3\ m_0$ for the aligned band and keep $m_B^* = m_0$ as a reference for the base band. The middle column (Fig. 2b) shows the peak PF in the case where only intra-band scattering is considered, as indicated by the orange transition arrows in Fig. 2a. The PF is improved even up to 100% in this case, as $\Delta E$ is reduced and the bands are aligned. This is expected, since a new, highly conducting band is now additionally contributing to transport, and the scattering rates are unaltered under intra-band scattering considerations. In contrast, under solemnly inter-band scattering conditions, the PF decreases upon band alignment, as shown in Fig. 2c. In this case the base valley is highly conducting (since we don't consider intra-valley scattering – the reason for the ultra-high PF values), and bringing another valley close-by introduces scattering for both, and reduces the conductivity and PF. Upon heavy valley band alignment, as shown in the second row, Fig. 2d-f, similar observations are encountered. Namely, under intra-band only conditions, improvements to the PF are realized, however, these are smaller compared to light band alignment (~ 20 %), since the heavy band offers reduced conductivity. Under inter-band only conditions the PF reduces in this case as well, since aligning increases scattering for both bands.

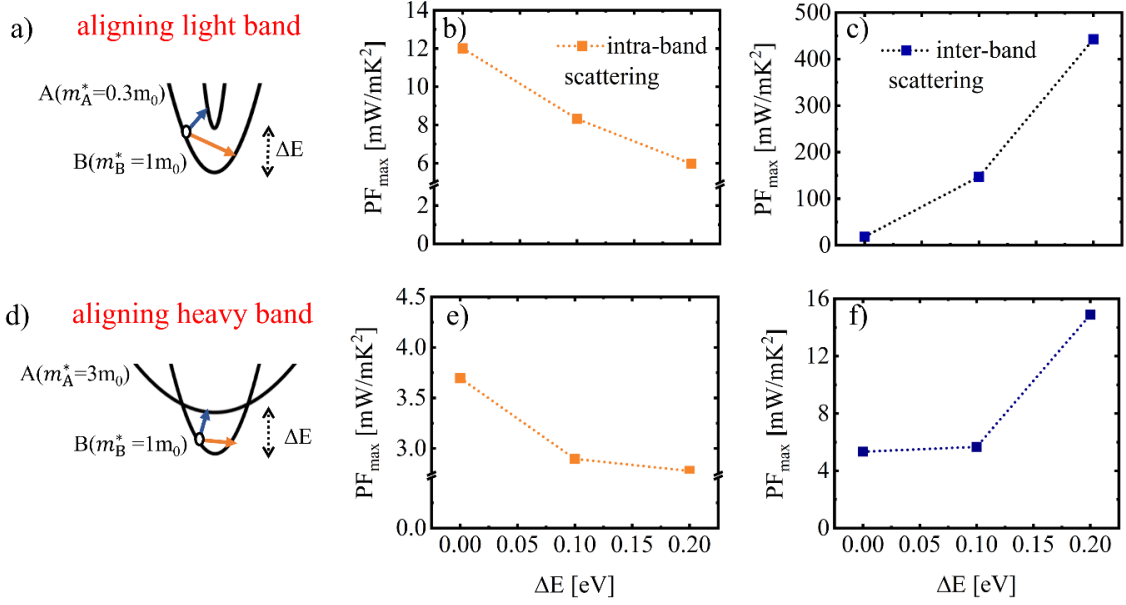

**Figure 2:** Upper row (a-c): The case for light band alignment. Lower row (d-f): The case for heavy band alignment. First column (a, d: illustrations of the intra-band (orange arrows) and inter-band (blue arrows) scattering processes in the bandstructures. Middle column (b, e): Maximum power factors for intra-band scattering for the two cases. Third column (c, f): Maximum power factors for inter-band scattering for the two cases.

These results illustrate that to realize PF improvements upon band alignment, intra-band scattering needs to dominate over the inter-band scattering processes, while benefits are more pronounced when a light band is brought into alignment. In general, intra-band processes are encountered when electrons scatter with acoustic phonons, anisotropic scattering mechanisms such as polar optical phonons and ionized impurities, and other static impurities like defects, boundaries, and alloy scattering [52]. Inter-valley processes, on the other hand, are typically dominated by large wave-vector optical phonon scattering. In typical TE materials the former are strong, thus it is more probable that intra-valley processes dominate, but only in the cases where the valleys of the different bands are placed farther away from each other in the Brillouin zone [11, 46, 50, 52, 56, 57].

*The degree of intra- compared to inter-band scattering:* In practice, independent intra- or inter-band scattering is not the typical case, but the study in Fig. 2 reflects the band alignment expectations upon these limiting conditions. In the typical scenario, both processes are present simultaneously, thus, below we quantify the degree upon which the intra-versus inter-band scattering strength ratio allows for PF improvements. For this, we vary the ratio of the deformation potentials which dictate the strength of each of the processes, $\frac{D_{\text{intra}}}{D_{\text{inter}}}$ from 2 (closer to intra-band scattering dominance), to 1, and then down to 0.5 (closer to mimicking inter-band scattering dominance).

We again start with the case of light band alignment, but we also consider different mass ratios, as shown in Fig 3a-c, from $m_A^* = 0.2$ m$_0$ (ultra-light) to $m_A^* = 0.8$ m$_0$ (closer to the base band mass of $m_B^* = 1$ m$_0$). In the case of the lighter aligned band mass, Fig. 3a, for the larger intra- versus inter-band deformation potential ratio ($\frac{D_{\text{intra}}}{D_{\text{inter}}} = 2$, purple line) a ~100% PF improvement upon full band alignment can be reached, as this also resembles the case in Fig. 2b for intra-band scattering dominance. As $\frac{D_{\text{intra}}}{D_{\text{inter}}}$ is reduced, and the situation moves closer to inter-band scattering dominance, the PF improvement is reduced as well. At equal deformation potentials the improvement drops to ~30%, whereas for $\frac{D_{\text{intra}}}{D_{\text{inter}}} < 1$ the PF is slightly degraded upon band alignment (black line) where strong inter-band scattering is highly supported by optical phonons. As the aligned band effective mass increases (Fig. 3b-c), but remains lower than the base band mass, the PF improvements at a specific $\frac{D_{\text{intra}}}{D_{\text{inter}}}$ and *ΔE* are reduced, while the degradation when inter-band becomes stronger, is larger (black lines). Essentially, the scattering introduced on carriers in the base band into each heavier-and-heavier aligned band, becomes stronger due to the ever-increasing scattering DOS, which limits its conductivity and the PF.

The trend of weakening the PF improvements at a certain $\frac{D_{\text{intra}}}{D_{\text{inter}}}$ ratio and ΔE continues as the aligned band mass, $m_A^*$, is increased and overpasses the mass of the base band ($m_B^* = 1$ m$_0$). This is shown in Fig. 3d-f, where now a heavy band is essentially brought into alignment, here with masses $m_A^* = 1.5, 2.5$ and 3.5 m$_0$. In this case, band alignment gives marginal PF benefits when the strength of intra-band scattering is larger, i.e. $\frac{D_\text{intra}}{D_\text{inter}} = 2$ (purple lines), but these benefits are diminished when the aligned band mass increases further (Fig. 3e-f). Here for any lower $\frac{D_\text{intra}}{D_\text{inter}}$ values the PF degrades for all masses of aligned bands. However, the PF increases somewhat, or at least does not decrease, upon band alignment for all the heavily aligned bands in limiting case of the presence of ultra-strong intra-valley scattering (gray line).

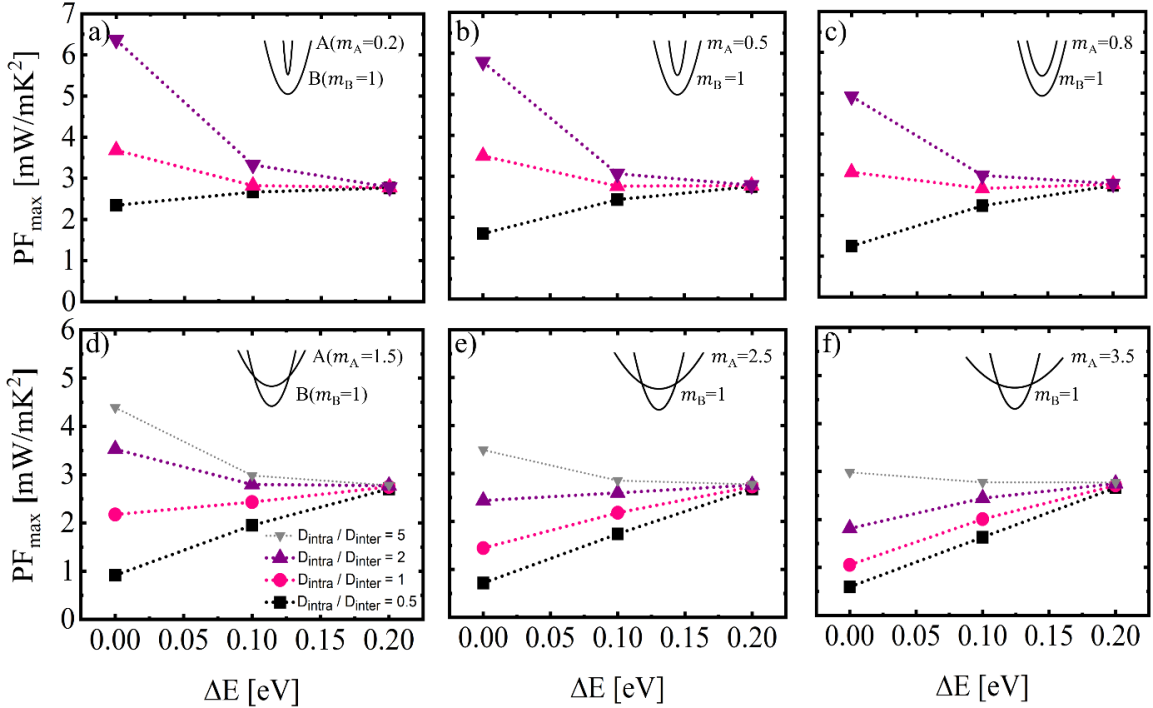

**Figure 3**: Maximum power factor upon band alignment of two single-valley bands versus their energy difference during the alignment process, $\Delta E$. The base band effective mass is set to $m_B^* = 1$ m$_0$ in all cases, while the effective mass of the aligned band ($m_A^*$) is noted in each subfigure in units of the free electron mass m$_0$. First row, (a-c): the case for a light aligned band. Second row (d-f): The case for a heavy aligned band. Sets of data for different intra- versus inter-band scattering cases ($\frac{D_\text{intra}}{D_\text{inter}}$) are shown. The PF improves upon full band alignment for larger intra-band scattering and lighter aligned bands.

Thus, this analysis shows that upon band alignment, PF benefits can be achieved when the aligned band mass is lighter than the base band mass. Even so, benefits are still reached when the relative scattering strength of the intra-band scattering process is of the order of, or stronger than the inter-band scattering process (as denoted here by the deformation potentials which determine the scattering process, i.e. $\frac{D_{\text{intra}}}{D_{\text{inter}}} \gtrsim 1$). As the mass of the aligned band becomes heavier, a much larger ratio is needed to realize benefits, which can be unrealistic in typical materials.

A summary of these findings is better illustrated in Fig. 4, in which presents a bar-chart of the percentage change of the PF upon full band alignment versus the mass ratio of the aligned and base band for the different $\frac{D_{\text{intra}}}{D_{\text{inter}}}$ ratios. For the large, aligned band effective mass ($m_A^*$) region (right side), most of the bars are negative, indicating reduction in the PF upon band alignment. As the aligned mass becomes smaller than the base band mass, only then the bars become positive and increase as that mass is reduced even further; and noticeably only for the cases where the intra-band processes are stronger compared to the inter-band scattering processes (purple and pink bars).

The red dashed-dot line and the square symbols indicate the case where *only intra-band scattering* is considered, i.e. the envelope of the best case scenario for the performance improvement upon band alignment. Note that as the mass of the aligned band is reduced to values smaller than $m_A^* < 0.5 m_0$ (left, upper part of the Fig. 4), we reach a point of diminishing returns, where the *PF improvement* upon band alignment starts to reduce. The absolute PF value is still larger as shown in Fig. S1.1 in the Supporting Information. However, since the much lighter aligned band contributes significantly to transport and the PF to begin with, aligning it with the heavier base band increases its scattering and offers less of an improvement. Interestingly, in this case, some inter-valley scattering can be beneficial to providing relative PF improvements, as in that case the lighter aligned band becomes more resistive and contributes to transport less to begin with (although the absolute value of the PF is reduced in this case).

With the black dashed-dot line we show the case where *only inter-band scattering* is considered, in this case it is the envelope for the worst performance improvements. This line does not follow the trend of the negative value bars, because those are not in the only inter-band scattering regime. The reason of this deviation is that for very low effective masses of the aligned band, that aligned band dominates the PF, which then is drastically reduced once it is brought in the vicinity of the base band which introduces scattering (similar to what we describe earlier in Fig. 2c).

An important point is which of the two quantities controls the behavior we show in Figs. 2, 3, and 4 above. As also shown and discussed for the case of Fig. 1, the conductivity and changes to that upon band alignment is what dominates the changes in the PF, while the Seebeck coefficient has a smaller influence. In the Supporting Information (Fig S5.1, and Fig S5.2) we show a large set of data for the TE coefficients in Fig. 2, 3 and 4, plotted versus the Fermi level. The fact that the conductivity determines the PF performance by a large degree is evident both in the case of light and heavy aligned bands, but also both in the cases of stronger intra-band scattering or stronger inter-band scattering. This reflects directly to our main observation, that light aligned bands of higher conductivity are what allow for higher PF improvements, rather than heavier aligned bands that offer higher density of states (DOS) but lower conductivity. Note that in our simulations we kept the base band constant and altered the aligned band. What matters finally is the ratio of the two masses, thus we could have altered the base mass and kept the aligned band mass the same, without changing our conclusions.

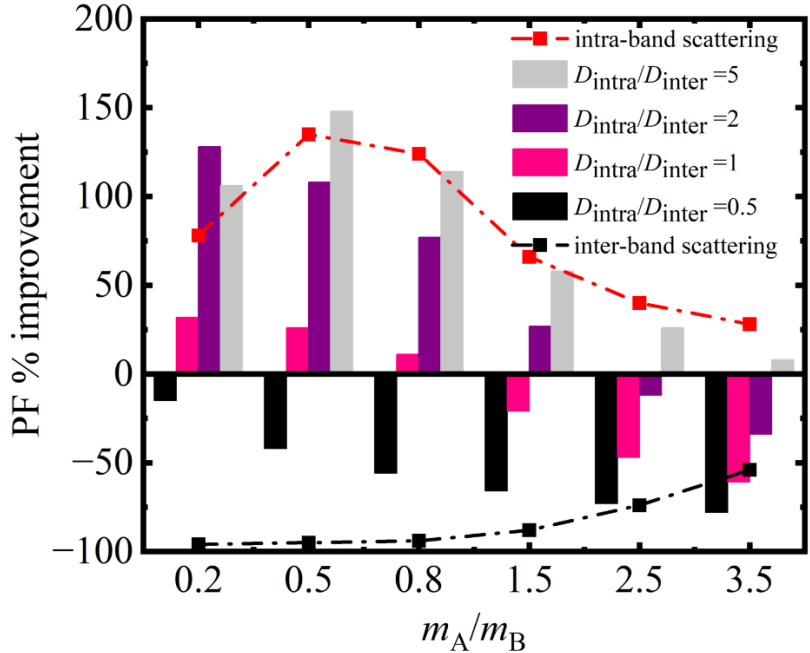

**Figure 4:** Bar chart illustrating the percentage improvement in the maximum PF upon full band alignment of two single-valley bands versus different effective mass ratios between the aligned and base bands, and for various intra- versus inter-band scattering strengths as set by the deformation potential ratios of the two processes, $\frac{D_{\text{intra}}}{D_{\text{inter}}}$. The red and black dashed-dotted lines show the cases of only intra- and only inter-band scattering, respectively.

The above analysis is performed for the simple case where a single degenerate band is aligned with another single degenerate band. In practice, typical materials involve bands at high symmetry points of the Brillouin zone with multiple band degeneracies. For example, the L valley of 8 half valleys ($N_v$=4), or the W valley of 24 one-third valleys ($N_v$=8). Thus, in practice we are faced with the possibility of aligning single, or multi-valley bands upon either single or multi-valley bands. We have identified that when the aligned band is heavier, and inter-band scattering dominates, benefits cannot be achieved. However, it is interesting to examine if this will still be the case if multi-valley bands are aligned, in which case multiple conduction channels are added to the transport energy window. This is discussed in the section below.

## IV. Band alignment with multi-valley bands

The schematics on the top panel of Fig. 5 illustrate the different scattering situations we consider. In all cases we consider the alignment of heavy effective mass bands, i.e. $m_A^* > m_B^*$. Since this was the case where it was most difficult to provide PF improvements, we seek to examine if there is a possibility to reverse this by aligning multiple bands. Here we consider intra-valley scattering for both base and aligned bands and inter-band scattering between them, but in addition we distinguish between two different scattering cases for the valleys of the same band: i) the case where inter-valley scattering is included (as indicated by the arrows in the top schematic), and ii) the case where inter-valley scattering is excluded (indicated by the magenta arrows in the lower row schematics). The latter is the case where ODP is weak, or the different valleys of the same band are placed far from each other in the Brillouin zone, and/or possibly IIS or POP is stronger, such that inter-valley scattering is the weaker of the scattering processes. In the former the reverse can be true. For simplicity we consider the same deformation potential across all scattering events, and same mass for the aligned band valleys ($m_A^*$=3$m_0$), a mass that was previously detrimental to the PF even at strong intra-valley scattering).

The first column (Fig. 5a-b) deals with the alignment of a multi-valley band with $N_v$=4, onto a single valley base band ($N_v$=1). This could be the case with TE materials such as half-Heusler (HH) alloys i.e. p-type TiCoSb, and TiNiSn [58,59] which have multi-valley heavy bands at L and a single valley band at the Γ point in the Brillouin zone (as shown in Fig S2.1 in the Supporting Information).

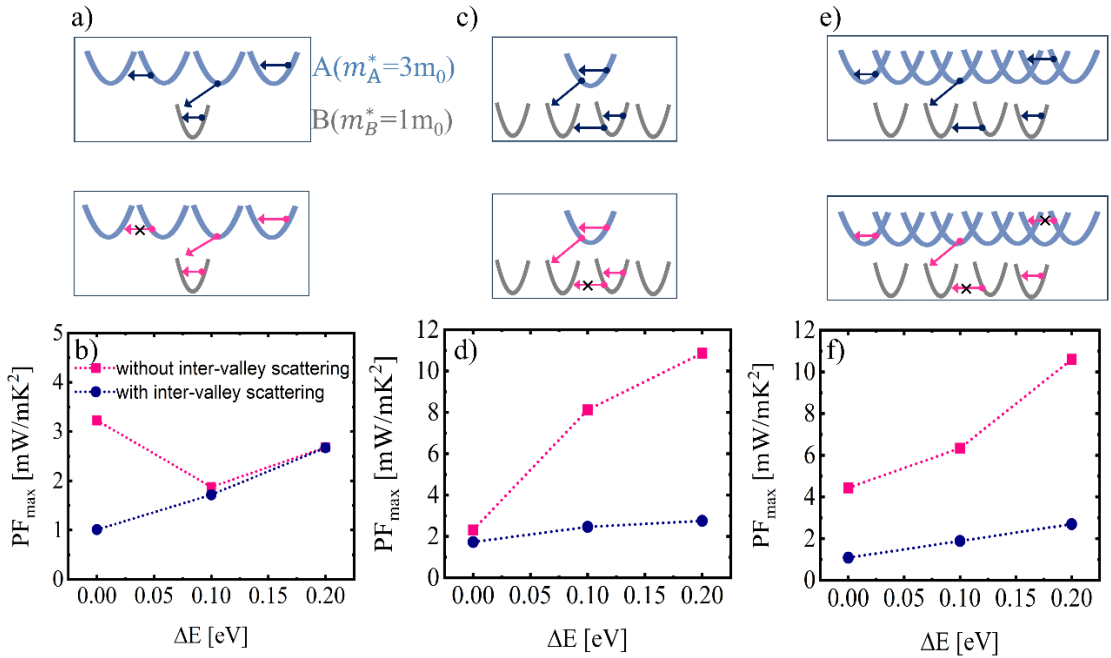

**Figure 5**: The maximum PF change upon band alignment of a band with heavier effective mass valleys compared to those of base band. Three different cases are shown as indicated by the schematics in the top rows: (a-b) A band with multiple heavy valleys is aligned with a lighter-valley base band. (c-d) A single-valley heavy band is aligned with a band of multiple lighter valleys. (e-f) A band with multiple heavier valleys is aligned with a band with multiple lighter valleys. The aligned band has valleys with $m_A^*$=3 $m_0$, whereas the base band has valleys with $m_B^*$=1$m_0$ (curvature not drawn to scale). Intra-valley and inter-band scattering are broadly considered. For each bandstructure two scattering cases are considered in addition: inter-valley scattering between valleys of the same band (blue lines), and the absence of such inter-valley scattering (magenta lines).

In the first scattering consideration case, where inter-valley scattering is allowed within the same band, band alignment degrades the PF performance (blue line in Fig. 5b). The reason is that since carriers from each valley scatter into each other, the contribution of that aligned band to transport is similar as that of a single band (more valleys, but more scattering). In addition, the inter-band scattering from the aligned to the base band that is present degrades the PF performance of the base band. Thus, the PF is overall degraded. In the second scattering case, where we exclude inter-valley scattering between the multi-valleys of the aligned band, the situation is slightly better for the PF upon band alignment (magenta line in Fig. 5b). Initially inter-band scattering for the lower energy band increases by four-fold (since $N_v^A = 4$) as the aligned band is lowered in energy, and the PF is lowered. This reduction is mitigated at full band alignment by the fact that four more valleys now contribute to transport. Still, however, even in this case, PF improvements are not observed, and at best the PF of the fully aligned bandstructure remains the same as the one of the fully non-aligned one.

Thus, aligning multiple (even heavy) valleys onto a single base valley might seem as a tempting favorable scenario, but it can be beneficial only when the combined system of those aligned valleys is more conductive compared to the base valley. For this, a large valley degeneracy is needed, but also there has to be reduced scattering within those aligned valleys themselves, i.e. reduced or no inter-valley scattering. This allows the high valley degeneracy to provide a highly conducting aligned system, despite their heavy mass. If inter-valley scattering is present, however, then the enhanced scattering within that aligned system reduces its conductivity and makes the situation similar to the alignment of a single heavy band, which as we discussed in the first section of the paper, it is not beneficial to the PF. We assume intra-band scattering is present in Fig. 5a, 5b. If intra-band scattering is strong, then the contribution to the PF of the base band is reduced due to scattering from the base band into the heavy valleys, thus this mitigates the PF improvements that can be reached. If intra-band scattering is not present, however, then the situation is beneficial to the PF as discussed in the first part of the paper. Essentially the base band conductivity remains intact, and additional conductivity from the heavy valley band system is brought into the picture, which increases the PF.

In the second case (middle column, Fig. 5c-d), we consider the reverse example, i.e. the alignment of a single-valley band (for example one that is located at the Γ point of the BZ), while the base band is now considered to consist of multi-valleys with $N_v^A = 4$ (at L). This could resemble materials such as p-type ZrCoBi, and ZrCoSb [60] where single-valley may be aligned with multi-valley band as shown in Fig. S2.2 in the Supporting Information. Here again, we consider intra-valley scattering for all valleys and intra-band scattering, and also as before: i) the case where inter-valley scattering is present between the base valleys, and ii) when inter-valley scattering is absent, as indicated by the arrows in the schematics on the top panel (Fig. 5c). In this case the four base valleys dominate transport to begin with. In the first case (blue line in Fig. 5d), inter-valley scattering keeps the PF low [32]. Aligning a heavier valley which introduces additional scattering, only degrades the PF more. In the second case, removing the inter-valley scattering in the base valleys to improve transport increases the amplitude of the PF to begin with (magenta line in Fig. 5d). However, upon full band alignment, the PF is reduced significantly due to the large increase in scattering that the large DOS of that aligned band introduces. Thus, band alignment does not help in this situation either.

In the third case (Fig. 5e-f), we consider aligning a multi-valley band ($N_v^A$=8, e.g. at the W point in BZ) to a multi-valley base band ($N_v^B$= 4, e.g. at the L point) as well. In general, the PF magnitude is greater in such cases, since many valleys offer more conduction channels, as also shown in a prior work [61]. Band alignment with these degeneracies may be realized in materials such as p-type NbFeSb, VFeSb and AgSbPbSnGeTe$_5$ as shown in Fig. S2.3 in the Supporting Information [62,63]. For example, the last case, high entropy material, gives high TE efficiency due to band convergence of four multiple valley bands i.e. L, W, Γ, and X. In such situations, the presence of a multi-valley base band provides a strong contribution in determining the PF to begin with, similar to the previous case. Thus, the maximum value of PF is observed before alignment (Δ*E*=0.2eV), also in agreement with a previous study [48]. In either case, in the presence of inter-valley scattering or its absence, the PF is reduced upon band alignment as shown in Fig. 5f (although the PFs are always larger as expected when inter-valley scattering is absent). This is because the weighted contribution to transport from the many heavy-band valleys is less in this case, since the base band also has multiple valleys, and the introduction of inter-band scattering degrades the PF.

Thus, the overall conclusion from this section, is that aligning heavier effective mass bands compared to the base band effective mass does not offer advantages to the PF in the presence of intra-/inter-band scattering, even if the aligned band consists of multiple valleys for transport, and even if inter-valley scattering within the same band is absent to increase the band's conductivity. At best, upon full band alignment, the PF retains its original value. One condition that can improve the PF upon full band alignment of a heavy band, is if that aligned band is highly conducting, for example if both the intra- and inter-valley scattering processes within that aligned band are weak, and degeneracy is high. This can be the case of aligning a multi-valley heavy band upon a single-valley base band as shown in Fig S3 in the Supporting Information. Then although upon band alignment the base band will experience increased scattering, the contribution of the highly conducting aligned band to transport can be significant, and the overall PF could increase. Such highly conducting high energy bands (compared to the lower energy base bands), however, would not be easily encountered in realistic material cases.

The situation for the PF is of course subject to change for lighter aligned band masses ($m_A^* \lesssim m_B^*$), and if inter-band scattering weakens on top of that, such that the aligned bands become more conductive this way. Such cases are in favor of PF improvements, as discussed in the previous section. Namely, the more the lighter valleys that are brought into alignment with respect to the number of heavier base valleys, the more the PF improvement. This is shown in Fig. 6, which again considers intra-valley and inter-band scattering, and the two cases of: i) inter-valley scattering (blue arrows in the top panel schematics), or ii) the absence of inter-valley scattering (magenta arrows in lower panel schematics). When multiple light bands are aligned on top of a single base band, PF improvements are reached (Fig. 6a-b). In this case the four additional bands bring almost four times the improvement the single band presented earlier (Fig. 1c), proving that valley degeneracy is an important parameter for high thermoelectric performance under these conditions [46]. If a single light-valley band (Fig. 6c-d) or a multiple light-valley band (Fig. 6e-f), is aligned on top of multiple heavy bands, no PF improvements are reached (magenta lines). This is because a multi-valley heavy base band offers a large inter-band scattering DOS as compared to single or multiple light aligned bands, in such a way, such that the aligned valleys cannot challenge the transport dominance of the multiple-valley heavy base bands. Only a moderate PF improvement can be reached in both cases, and this only when the conductivity of the base heavy band is already weakened by inter-valley scattering (blue line). Thus, overall, even in this case of lighter aligned bands, the aligned band needs to provide significant conductance benefits over the base heavier band valleys for improvements to be achieved. PF improvements can be reached in this case, by even up to 100% and beyond, i.e. more than doubling the PF, but as in most case described above when: i) intra-band scattering dominates over inter-band scattering, (smaller improvements are obtained when inter-band scattering exists, even down to the level when intra- and inter-band scattering deformation potentials are comparable to each other ($\frac{D_{\text{intra}}}{D_{\text{inter}}} \gtrapprox 1$), and ii) when aligning highly conducting (degenerate) valleys for which not only inter-band scattering is weak (between he aligned band and the base band), but also when inter-valley scattering is also weak (between the degenerate valleys themselves).

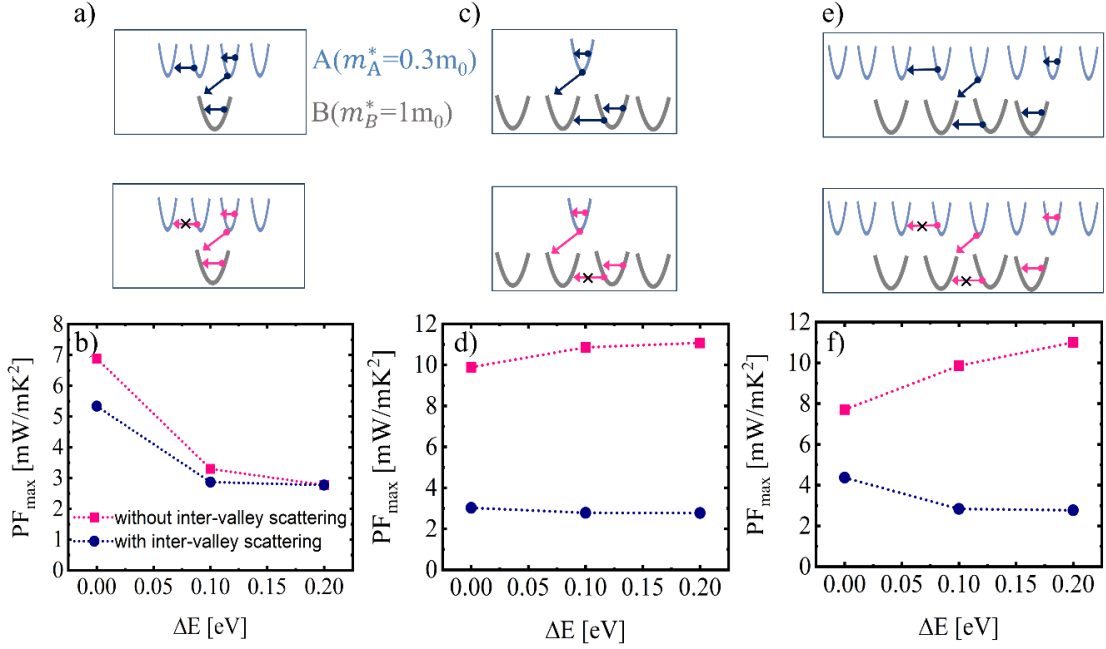

**Figure 6**: The maximum PF change upon band alignment of band with lighter effective mass valleys compared to those of base band. Three different cases are shown as indicated by the schematics in the top row: (a-b) A band with multiple light valleys is aligned with a single-valley heavier mass base band. (c-d) A single light valley band is aligned with a base band with multiple heavier valleys. (e-f) A band with multiple light valleys is aligned with a multiple-valley heavier band. The aligned band has valleys with $m_A^* = 0.3$ $m_0$ whereas the base band has valleys with $m_B^*=1$ $m_0$ (curvature not drawn to scale). Intra-valley and inter-band scatterings are broadly considered. For each bandstructure two scattering cases are considered in addition: inter-valley scattering between valleys of the same band (blue lines), and the absence of such inter-valley scattering (magenta lines).

## V. Conclusion

In this work, we used a multi-band parabolic effective mass model and the Boltzmann Transport Equation to investigate and quantify the effect of band alignment on the thermoelectric power factor of multi-band/valley electronic structure materials. We have considered different band alignment conditions that can improve the power factor in the presence of combinations of intra- and inter-band/valley scattering. We show that in general, upon band alignment, power factor improvements can be reached when the aligned band is highly conducting with respect to the base band, and it does not interact strongly in terms of inter-band scattering between them. Specifically, this can be achieved when intra-band scattering dominates over inter-band scattering, and primarily when a lighter valley band is brought into alignment, while smaller improvements are realized when a heavy valley band is brought into alignment. In practical terms this can reach around 100% improvements.

In the presence of dominating inter-band scattering, power factor improvements cannot in general be achieved, especially when heavy-valley bands are brought into alignment. Even when the heavy-valley aligned band consists of multiple degenerate valleys, benefits still cannot be realized unless that heavy-valley band is (unrealistically) highly conducting with limited intra- and inter-valley scattering between itself and it is not interacting strongly with the base band. Note however, that many promising thermoelectric materials are polar, and ionized impurity scattering is strong under high doping conditions. These are both anisotropic scattering mechanisms which favor intra-valley scattering. In these materials the dominance of intra-valley scattering can allow power factor benefits upon-band alignment. Our findings would help to correctly and easily identify promising material candidates to focus band alignment studies on, which would expedite the realization of high-performance thermoelectric materials.

# Supporting Information:

The Supporting information file contains information on:

1. Power factor values for the data in Fig. 4 of the main paper

2. Examples of real material bandstructures with misaligned bands

3. Heavy band alignment upon only inter-band/valley scattering considerations

4. Seebeck coefficient shape explanations

5. Full data for Figs 2 and 3 of the main paper

6. Results on altering the base band effective mass

# Acknowledgements:

This work has received funding from the UK Research and Innovation fund (project reference EP/X02346X/1). SEAA acknowledges Bhawna Sahni, Rajeev Dutt, and Pankaj Priyadarshi for helpful discussions.

# Supporting Information

## Conditions for thermoelectric power factor improvements upon band alignment in complex bandstructure materials

Saff E Awal Akhtar[+] and Neophytos Neophytou[*]

[+]saff-e-awal.akhtar@warwick.ac.uk, [*]N.Neophytou@warwick.ac.uk

## Section 1: Power factor values for the data in Fig. 4 of the main paper

The figure below presents the power factor (PF) for the data presented in Fig. 4 of the main paper, i.e. as the mass of the aligned band changes from a light mass (left side, $m_A^* < 1m_0$) to a heavy mass (right side, $m_A^* > 1m_0$)) while the base band has $m_B^* = 1m_0$. The figure shows the actual PF values, but with positive bars it shows the power factor cases that result in improvement of the PF upon band alignment, while with negative bars indicates the cases where full alignment results in PF degradation. The actual PF values are the absolute values.

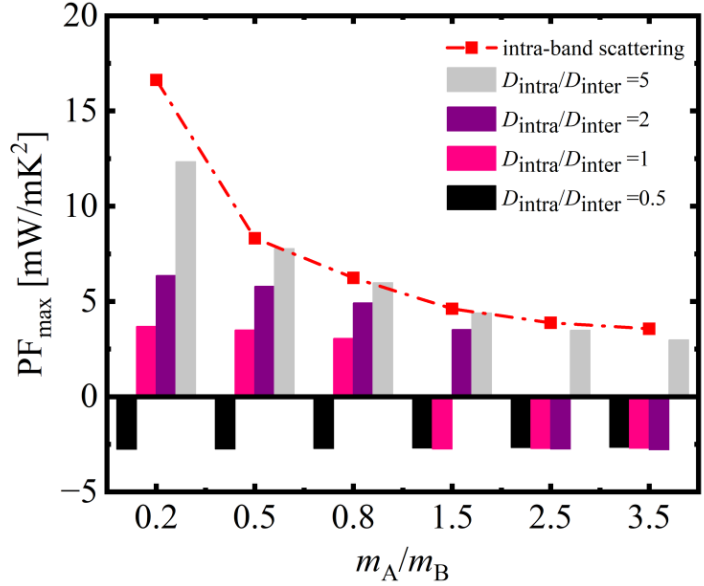

**Figure S1.1**: Bar chart illustrating the maximum PF during the band alignment process for the cases described in Fig. 4 of the main text. The aligned/base bands are initially separated by $\Delta E = 0.2$ eV, while upon alignment they have $\Delta E = 0$ eV. Values for different effective mass ratios are shown between the aligned ($m_A^*$) and base ($m_B^*$) bands. These calculations have been performed for various intra- versus inter-band scattering strengths as set by the deformation potential ratios of the two processes, $\frac{D_{\text{intra}}}{D_{\text{inter}}}$ and noted in the figure. The red dashed-dotted line shows the cases of only intra-band scattering considerations.

## Section 2: Examples of material bandstructures with misaligned bands

Here we show the relevant bandstructures of half Heusler thermoelectric materials indicating the possibility of band alignment in the valence band for TiCoSb, TiNiSn (Fig. S2.1-aligning many valleys from L onto a single Γ valley), ZrCoBi, and ZrCoSb (Fig. S2.2-aligning a single Γ valley onto multiple L bands), and VFeSb, NbFeSb (Fig. S2.3-aligning mutliple W bands onto multiple L bands). The dots show the band extrema. The red arrows show the possible bands to be aligned with the base valence band. ΔE(eV) in each figure represents the energy offset between the base and aligned bands in the valence band.

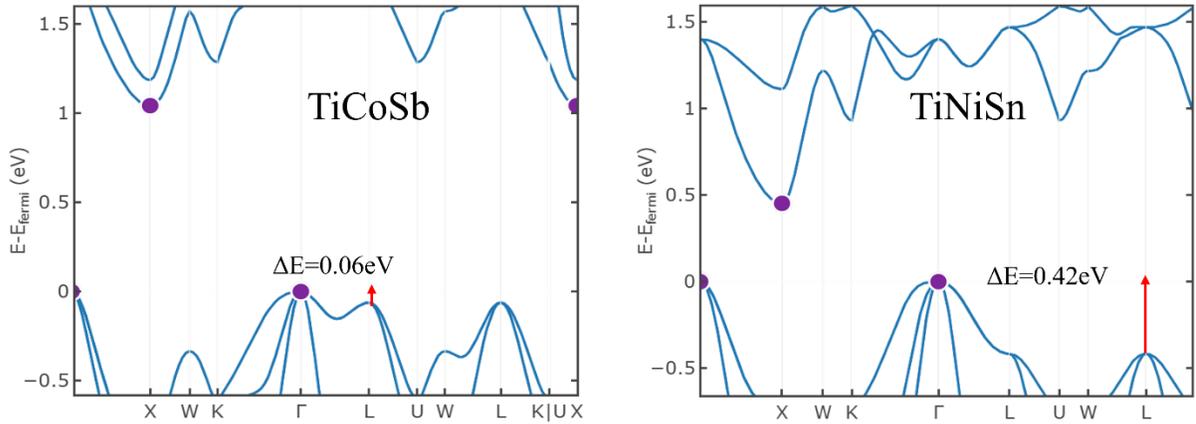

**Figure S2.1**: The bandstructures of (a) TiCoSb [1], and (b) TiNiSn [1] around the bandgap. These are examples of multi-valley band (at L) alignment upon single-valley band (at Γ).

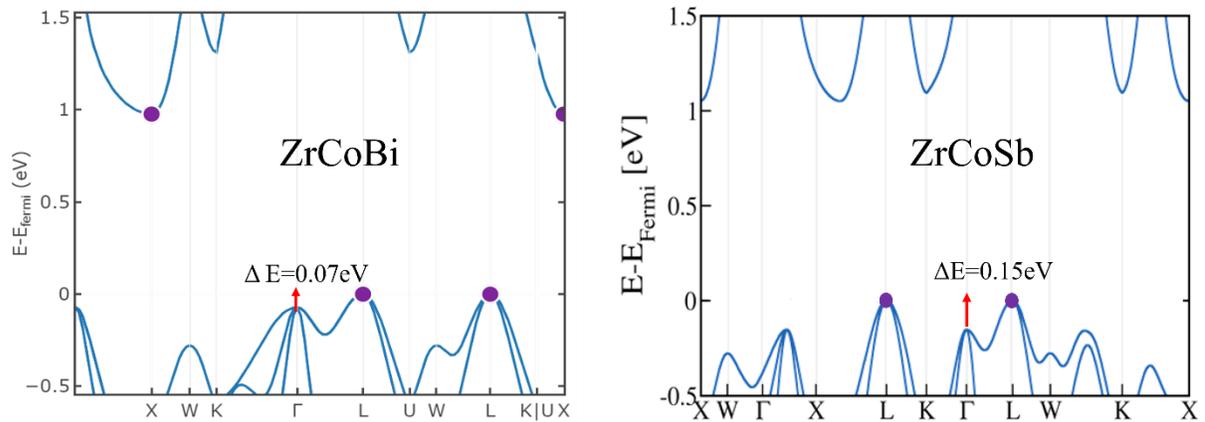

**Figure S2.2**: The bandstructures of (a) ZrCoBi [1], and (b) ZrCoSb [self-calculated] around the bandgap. These are examples of single-valley band (at Γ) alignment upon multi-valley band (at L).

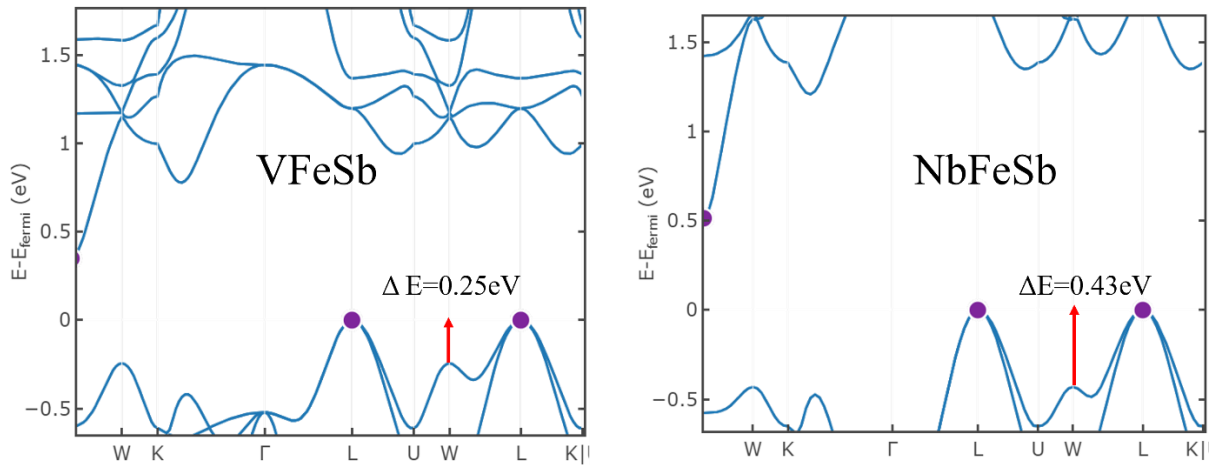

**Figure S2.3**: The bandstructures of (a) VFeSb [1], and (b) NbFeSb [1] around the bandgap. These are examples of multi-valley band (at W) alignment upon multi-valley band (at L).

## Section 3: Heavy band alignment upon only inter-band/valley scattering considerations

Figure S3 shows a schematic illustration of inter-band/valley scattering only (exclude intra-band/valley scattering) for multi-valley heavy band alignment upon a single band. When only inter-band scattering is present (without inter-valley), then power factor benefit is achieved upon full band alignment (green arrows and lines). In the case where inter-valley scattering is present, i.e. the aligned bands become now more resistive, PF degradation is experienced upon full band alignment.

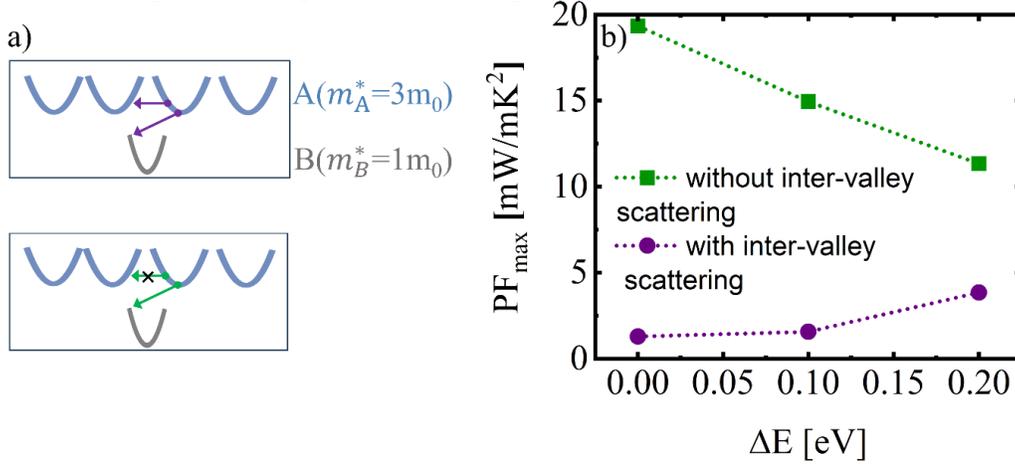

**Figure S3**: Maximum PF for two cases upon band alignment. Heavy band alignment between aligned band A ($m_A^*$=3) with valley degeneracy $N_v^A = 4$, and base band B ($m_B^*$=1). With green we show the case where only inter-band/valley scattering is present, whereas with purple we also consider inter-valley scattering between the valleys of multi-valley bands as well (see arrows in band illustrations). In both cases intra-valley scattering is omitted.

## Section 4: Seebeck coefficient shape explanations

Figure S4.1 shows the Seebeck coefficients from Fig. 1 of the main paper in zoomed-in versions (same as the insets of Fig. 1) to illustrate how they change in the higher Fermi level regions where the PF peaks, upon bringing in another valley. Changes in the Seebeck coefficient upon band alignment are typically hidden when we plot the Seebeck coefficient from different scenarios in terms of the Fermi level, $E_F$. These changes are much more pronounced if the Seebeck coefficients are plotted versus density, $n$. Thus, we also present the figures for $S$ plotted versus density, where the changes are more pronounced and follow the usual trend of increasing Seebeck with increasing density of states, DOS.

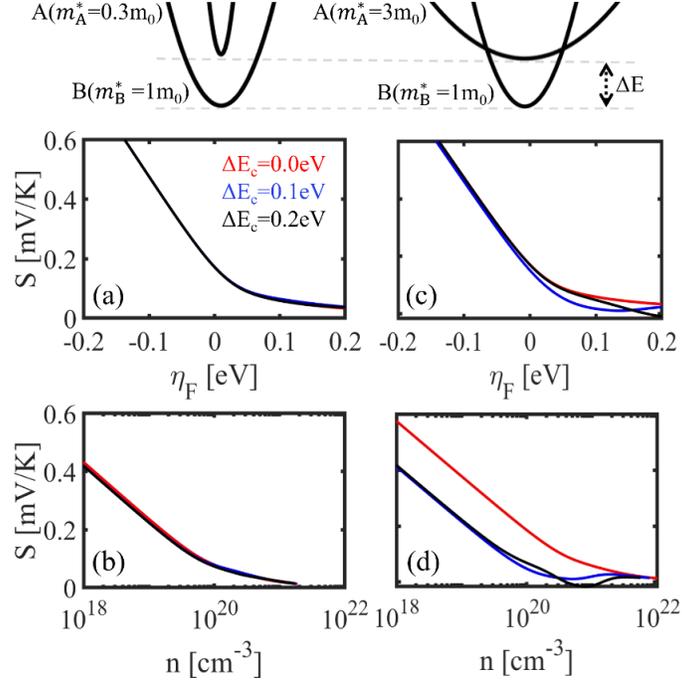

**Figure S4.1:** The Seebeck coefficients from Fig. 1 of the main paper in zoomed-in versions plotted versus Fermi level (middle column), and versus density (third column).

To explain the almost negligible shift in S when plotted versus the Fermi level, we start with the definition of the Seebeck coefficient as 'the average energy of the current flow with respect to the Fermi level'. To see this, we can rearrange the original equation for the Seebeck coefficient:

$$S_{(E_F)} = \frac{qk_B}{\sigma_{(E_F)}} \int_E G(E) \left(\frac{E-E_F}{k_B T}\right) \left(-\frac{\partial f}{\partial E}\right) dE \quad (1)$$

where $G(E) = v(E)^2 g(E) \tau(E)$ is the transport distribution function, and $\left(-\frac{\partial f}{\partial E}\right)$ is the energy derivative of Fermi Dirac distribution function, sharply peaked around Fermi level.

We can expand the numerator of Eq. (1) as:

$$\int_E G(E)(E - E_F)\left(-\frac{\partial f}{\partial E}\right) dE = \int_E G(E)(E)\left(-\frac{\partial f}{\partial E}\right) dE - E_F \int_E G(E)\left(-\frac{\partial f}{\partial E}\right) dE$$

(2)

Substituting back into the main equation (Eq. 1), we obtain:

$$S_{(E_F)} = \frac{q}{T} \frac{\int_E G(E)(E)\left(-\frac{\partial f}{\partial E}\right) dE - E_F \int_E G(E)\left(-\frac{\partial f}{\partial E}\right) dE}{\sigma_{(E_F)}}$$

(3)

and with

$$\sigma_{(E_F)} = \int_E G(E)\left(-\frac{\partial f}{\partial E}\right) dE,$$

(4)

we obtain:

$$S_{(E_F)} = \frac{q}{T} \frac{\int_E G(E)(E)\left(-\frac{\partial f}{\partial E}\right) dE}{\int_E G(E)\left(-\frac{\partial f}{\partial E}\right) dE} - \frac{E_F \int_E G(E)\left(-\frac{\partial f}{\partial E}\right) dE}{\int_E G(E)\left(-\frac{\partial f}{\partial E}\right) dE} \qquad (5)$$

Here the first term is the 'weighted energy of conducting carriers that participate in electrical transport, or the so-called 'energy of the current flow', ⟨E⟩, as:

$$\langle E \rangle = \frac{\int_E G(E)(E)\left(-\frac{\partial f}{\partial E}\right) dE}{\int_E G(E)\left(-\frac{\partial f}{\partial E}\right) dE}$$

(6)

Finally we can define the Seebeck coefficient in form of the energy of the current flow, ⟨E⟩, as:

$$S_{(E_F)} = \frac{q}{T} [\langle E \rangle - E_F] \qquad (7)$$

Thus, *S* is defined by the energy of the current flow with respect to the Fermi level. We can now imagine the scenario where we align a second valley on the base valley of equal DOS. This will double the DOS at the band edge. If we keep the Fermi level fixed, at the original position (as a gedanken experiment at this point – essentially allowing the density to double), clearly *S* will not change, as the energy of the current flow in each valley independently, and combined, will be the same as the one in the original base valley. This is the case when we plot *S* versus $\eta_F$, as in this work. If the mass of the aligned band varies, then some small changes in *S* are expected as the energy of the current flow will be changed slightly, however, not due to the DOS directly, but due to the complexities of transport. In reality, however, if the carrier density in the material remains constant (i.e. as determined by the doping density) – and this can be a more realistic scenario, then when the DOS doubles the Fermi level will shift lower compared to the band edge to retain the carrier density. In this case the Seebeck coefficient increases, since ⟨E⟩ -$E_F$ increases, essentially shifting the *S* vs *n* curve to the right (see red line in the figure above). Note that this is exactly what happens in the Pisarenko plot when a material acquires a larger DOS upon alloying or other band modifications, at (mostly) unchanged density. Of course, how we plot the Seebeck coefficient does not affect our results, as the conductivity and PF will also shift together with *S*, and the peak PF that the paper examines, will be unaffected. This is shown clearly in the Fig. S4.2 below, which shows that all TE coefficients shift once plotted versus density.

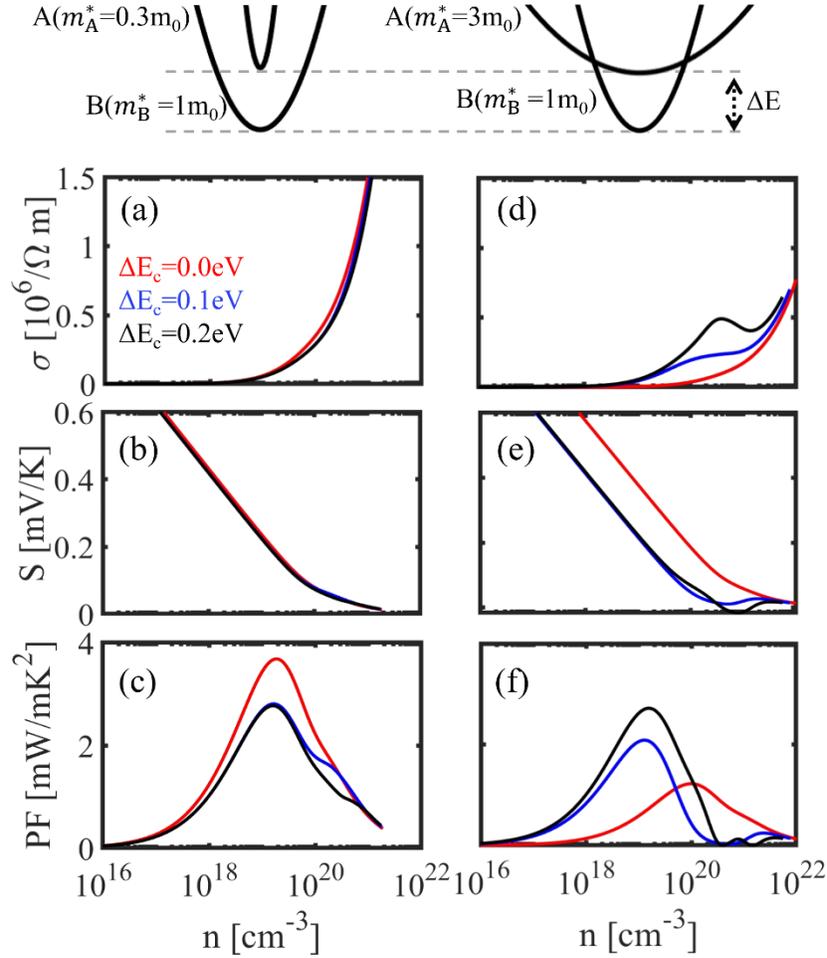

**Figure S4.2:** The TE coefficients for Fig. 1 of the main manuscript plotted versus density.

**Section 5: Full data for Figures 2 and 3 of the main paper**

Here we show full data for $\sigma$, $S$ and PF for what presented in Figs. 2, 3 and 4 in the main paper. Figure S5. 1 shows in each of the sub-figures two sets of three lines (for $\Delta E = 0$ eV, $\Delta E = 0.1$ eV, and $\Delta E = 0.2$ eV): the first set is for intra-band scattering only (dashed lines) and the second set for inter-band scattering only (solid lines). In the case of a light aligned band (left column), in the intra-band scattering only case (dashed lines) the improvements in the PF originate from improvements in $\sigma$, whereas $S$ does not change much, as also explain in the response to the first comment. In the case of inter-band scattering only (solid lines), large $\sigma$ and PF values are observed in the misaligned case (black solid lines) since intra-band scattering is suppressed, and the conductivity of the lower base band continues to increase until the upper band is reached. Thus, all the TE coefficients are shifted towards higher energies, and the higher conductivity leads to very-high PFs (although this is under the unrealistic assumption of complete suppressed intra-band scattering). As the bands are aligned, a left shift in the TE coefficients is observed, and a strong reduction in the PF due to increase in inter-band scattering as the upper band aligns with the base band.

In the case of Fig. 3b and 3e, the full data is shown in Fig. S5.2 below. In this case we also show two sets of three lines (for $\Delta E = 0$ eV, $\Delta E = 0.1$ eV, and $\Delta E = 0.2$ eV): the first set is for stronger intra-band scattering (dashed lines), whereas the second set is for stronger inter-band scattering (solid lines). In the case of a light aligned band (left column), what contributes to the changes in the PF are primarily changes in the conductivity, whereas the Seebeck coefficient shows less variations (some variations at high Fermi levels are observed as indicated in the zoomed-in inset). In the case where we consider stronger intra-band scattering (dashed lines), an increase in the conductivity upon band alignment is observed, which translates to PF improvements, whereas in this case the Seebeck coefficient slightly decreases upon band alignment (at high Fermi levels – see inset). In the case where we consider stronger inter-band scattering (solid lines), PF reduction is observed upon band alignment, and this reduction originates again from the conductivity reductions (compare the black to the fully aligned red line). Following the expected inverse trend, in this case the Seebeck coefficient increases slightly upon full alignment, but not enough to result to PF improvements (see higher red line in inset). In the second case (right column), when considering heavy band alignment, upon stronger intra-band scattering (dashed lines) the PF is degraded only slightly upon band alignment. In this case neither the conductivity nor the Seebeck coefficient experience noticeable changes. In the case of stronger inter-band scattering, however, the PF is strongly reduced upon band alignment (solid lines – red is lower than the black one). Again, in this case it is the conductivity which is reduced noticeably. The Seebeck coefficient increases slightly (inset - compare the black line to the fully aligned case shown by the red line), but still the conductivity is what dominated the PF reduction.

So overall, the changes in the PF are controlled by changes in the conductivity, rather that the Seebeck coefficient.

We have to stress, however, that upon performing computational studies, one has to choose what quantity will be represented in the x-axis, and that has consequences on the phenomenological shifts of the quantities on the x-axis, but not on the PF shape or the PF peak, other than a shift in the x-axis of the plots.

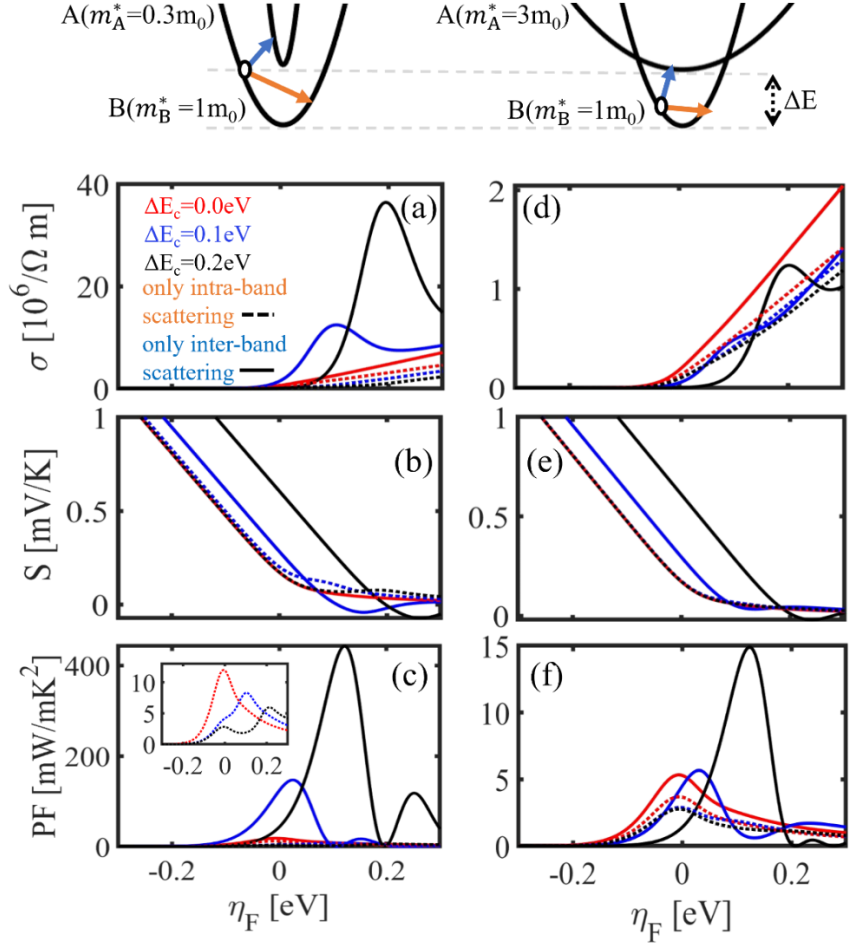

**Figure S5.1:** Full data lines for Fig. 2 of the main manuscript. Showing $\sigma$, $S$ and PF for light and heavy band alignment.

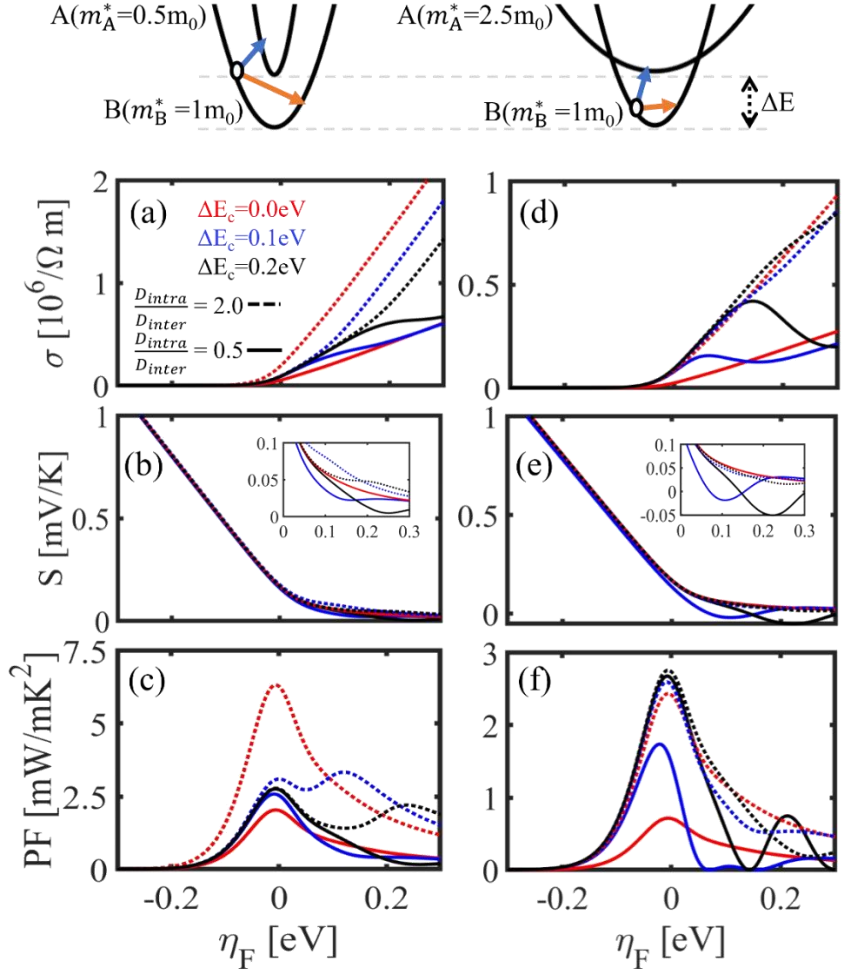

**Figure S5.2:** Full data lines for Fig. 3b (left column) and Fig. 3e (right column) of the main manuscript.

### Section 6: Altering the base band effective mass

In our simulations we kept the mass of the base valley at 1 $m_0$ at all cases to keep the number of possible parameters that we vary at a minimum. What matters finally in our conclusions is the relative mass variation between the base valley effective mass and the aligned valley effective mass. Thus, if we lower the effective mass of the base band, we will require to align even lighter upper valley to achieve the same PF improvement ratio. With regards to our overall conclusion, that the composite valley system we align needs to be of higher conductivity compared to the base valley, it follows again that a lighter base band will require a more conductive aligned valley system for PF improvements, whereas with a heavier base mass we can afford to align less conducting valleys.

Below we illustrate this with relevant simulations, in which we keep the aligned band mass constant as shown in Fig. S6 and vary the base band mass. In the first case (first row) we start with the case presented in Fig. 1a in the main paper, where a PF increase of 33% is observed for $m_B = 1\ m_0$. We then increase the mass of the base band. As the base band mass increases the overall PF amplitude decreases since now the heavy base band degrades conductivity. Upon aligning the upper light band in these heavier base band cases, a similar improvement is observed. This is expected. At band alignment the light band will tend to provide a certain boost in the PF on top of the heavy base band in absolute terms. But scattering into the heavy band increases as the base band mass increases and the conductivity of the aligned band reduces in absolute terms. On the other hand, the conductivity of the base band and its PF is also reduced in absolute terms. Thus, there is a lower PF starting point in the case of heavy base bands, and a lower PF increase from the light aligned band contribution, thus overall, the relative PF increase remains similar when we change the base band mass. This pretty much follows the pink line in Fig. 4 of the main paper (in the left half of the figure with $m_A/m_B$ from 0.3 down to 0.15), where the PF improvements tend to saturate with the $m_A/m_B$ ratio.

In the second case (second row), we consider a heavy mass aligned band, which we keep constant, and increase the mass of the base band. The most left case is the one presented in Fig. 1b of the main paper, where a 55% reduction in the PF is observed (aligning a heavy band). As the base band mass increases, the overall PF amplitude decreases since now the heavy base band degrades conductivity. The relative PF reduction from the mis-aligned to the fully aligned case however, is reduced in percentage. This is also expected. The black line, mis-aligned case is dominated by the lower base band for which the PF is reduced as the base band mass increases. Once a heavy band is aligned with the light base band in the first left-most case, the conductivity of the base band suffers significantly (large relative additional scattering rate introduced), and a larger PF reduction is observed. As the base band mass increases, its conductivity suffers relatively less (i.e. the additional scattering rate introduced is relatively less and less as the base band becomes heavier), and the PF is degraded less in relative terms. This is pretty much following the pink line in Fig. 4 of the main paper (in the right half of the figure with $m_A/m_B$ from 3 down to 1.5).

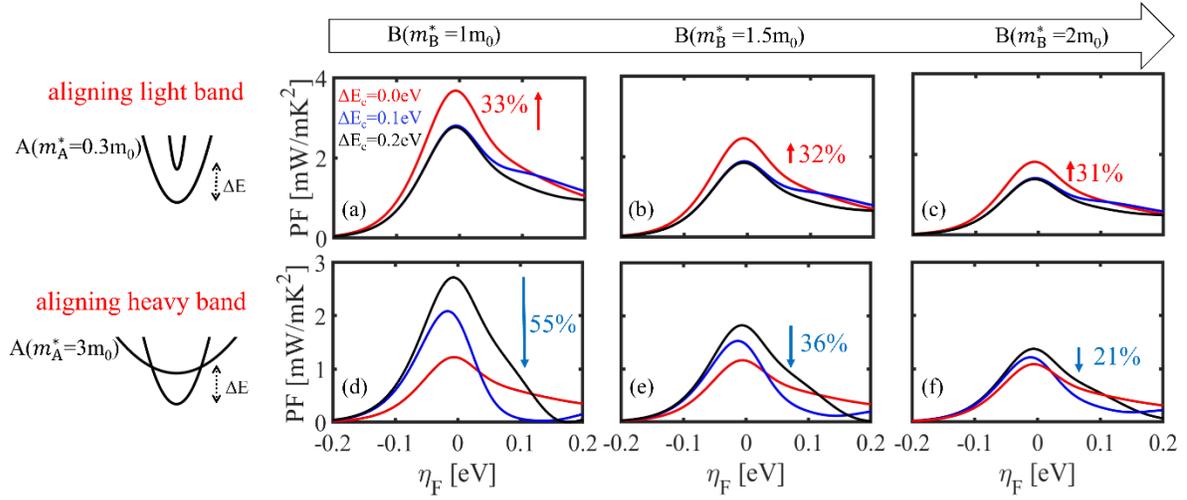

**Figure S6.1:** Simulations considering changes in the base band mass, rather than the aligned band mass.

References
[1] Jain, A.; Ong, S.P.; Hautier, G.; Chen, W.; Richards, W. D.; Dacek, S.; Cholia, S.; Gunter, D.; Skinner, D.; Ceder, G.; Persson, K. A. Commentary: The Materials Project: A materials genome approach to accelerating materials innovation. *APL Mater.* **2013**, 1(1).